\documentclass[aps,pra,english,twocolumn,showpacs,preprintnumbers,amsmath,amssymb,floatfix,superscriptaddress]{revtex4-1}

\usepackage[T1]{fontenc}
\usepackage[latin9]{inputenc}
\usepackage{graphicx}
\usepackage{amssymb}
\usepackage{babel}


\usepackage{babel}
\makeatother

\newcommand{\be}{\begin{equation}}
\newcommand{\ee}{\end{equation}}

\begin{document}

\title{Sign of the Casimir-Polder interaction between atoms and oil-water
   interfaces: Subtle dependence on dielectric properties}

\author{Mathias Bostr{\"o}m}
\email{mabos@ifm.liu.se} 

\affiliation{Department of Energy and Process Engineering, Norwegian University of Science and Technology, N-7491 Trondheim, Norway}
\affiliation{Division of Theory and Modeling, Department of Physics, Chemistry and Biology, Link\"{o}ping University, SE-581 83 Link\"{o}ping, Sweden}
\affiliation{Department of Applied Mathematics, Australian National University, Canberra, Australia}

\author{Simen {\AA}. Ellingsen}

\author{Iver Brevik}
\affiliation{Department of Energy and Process Engineering, Norwegian University of Science and Technology, N-7491 Trondheim, Norway}

\author{Drew F. Parsons}
\email{Drew.Parsons@anu.edu.au}
\affiliation{Department of Applied Mathematics, Australian National University, Canberra, Australia}

\author{Bo E. Sernelius}
\email{bos@ifm.liu.se}
\affiliation{Division of Theory and Modeling, Department of Physics, 
Chemistry and Biology, Link\"{o}ping University, SE-581 83 Link\"{o}ping, Sweden}

\begin{abstract}
We demonstrate that Casimir-Polder energies between noble gas atoms
(dissolved in water) and oil-water interfaces are highly surface specific. Both repulsion (e.g.\ hexane) and
attraction (e.g.\ glycerine and cyclodecane) is found with different oils. For several intermediate oils (e.g.\ hexadecane, decane, and cyclohexane) both
attraction and repulsion can be found in the same system. Near these oil-water interfaces the
interaction is repulsive in the non-retarded limit and turns attractive at larger distances as
retardation becomes important. These highly surface specific interactions may have a role to play
in biological systems where the surface may be more or less accessible to dissolved atoms.
\end{abstract}


\maketitle

The Casimir--Polder interaction \cite{CasPold} between polarizable particles 
and
a wall has received intense attention in 
recent decades.
The theoretical framework for this interaction was worked out a long time ago \cite{CasPold,london30,Lifshitz,Dzya,barton70}, yet it remains a topic of great interest due to its many applications in biological, chemical and atomic systems. For reviews, cf.,\ e.g.,\ \cite{Ser,buhmann07,scheel08}.

An interesting aspect of the Casimir--Lifshitz and Casimir--Polder forces is that according to theories these forces can either be attractive or repulsive.  
Anderson and Sabiski  performed experiments on films of liquid helium on calcium fluorite, and similar molecularly smooth surfaces \,\cite{AndSab}. The film thicknesses ranged from 10-200 \AA \,\cite{AndSab}.  In these measurements the repulsive van der Waals potential opposed the gravitational potential \,\cite{AndSab}. A  good agreement was found\,\cite{Rich1} between experimental data and  Lifshitz theory \cite{Lifshitz}. In a set of experiments that inspired the present work, Hauxwell and Ottewill\,\cite{Haux} measured the thickness of films of oil on water near the alkane saturated vapor pressure, an asymmetric system (oil-water-air) in which the surfaces are molecularly smooth. For this system $n$-alkanes up to octane spread on water. Higher alkanes do not spread. The phenomenon depends on a balance of van der Waals forces against the vapor pressure\,\cite{Rich, Haux}. The net force, as a function of film thickness depends on the dielectric properties of the oils \,\cite{Masuda}. As demonstrated\,\cite{Rich} it involves an intricate balance of repulsive and attractive components from different frequency regimes. When the ultraviolet components are exponentially damped by retardation, the opposing (repulsive) infrared and visible components take over\,\cite{Maha, Maha2}. Other studies discussing the transition between attractive and repulsive interactions are found in Refs.\,\cite{Elb, Wil, Lev, Mag}.

\begin{figure}
\includegraphics[width=8cm]{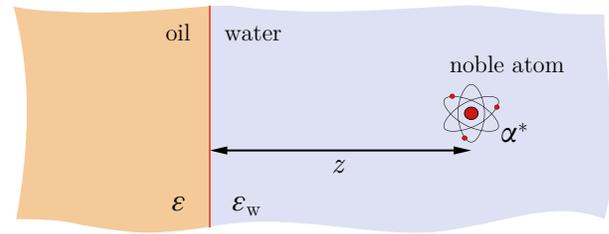}
\caption{(Color online) Geometry considered: an atom immersed in water near an interface with an oil}
\label{figu1}
\end{figure}
Surfaces of interest in biology and biotechnology may involve alkane molecules creating an oil-water interface. In this Brief Report we will demonstrate that the Casimir-Polder interaction between dissolved atoms and different oil-water interfaces may be either repulsive or attractive, or as we will se for some of the alkanes, it may change from repulsion to attraction as the distance to the interface increases. The relevant geometry is sketched in Fig.\,\ref{figu1}. The net dispersion forceresults from a delicate balance of attractive and repulsive frequency contributions and the net sign for a specific oil and specific distance can change from one atom to another.  
In this manner
the interface selects which noble atoms will be able to approach it, and which will not. A theoretical determination of whether a given atom will be attracted or repelled by an interface to a given dielectric material can thus only be made upon a careful study of the dielectric properties of water, atom and dielectric in combination, over the full frequency range. 

In the following we briefly review the governing equations for dispersion interaction between an atom and an interface and present a series of numerical examples which are discussed, whereupon our conclusions are drawn and presented.

\begin{figure}
\includegraphics[width=\columnwidth]{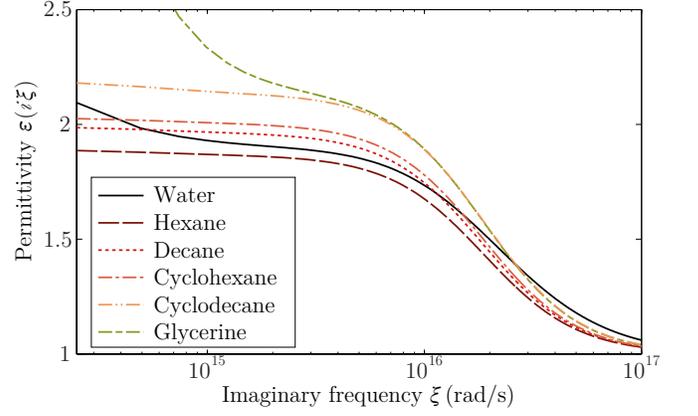}
\caption{(Color online) Dielectric functions of water \,\cite{Ser}, hexane, decane, cyclohexane, cyclodecane, and glycerine \,\cite{Masuda} for imaginary frequencies. Near the quasistatic limit the permittivity of water rises to $77.9$ whereas the same limit for glycerine is substantially lower.}
\label{figu2}
\end{figure}

\begin{figure}
\includegraphics[width=\columnwidth]{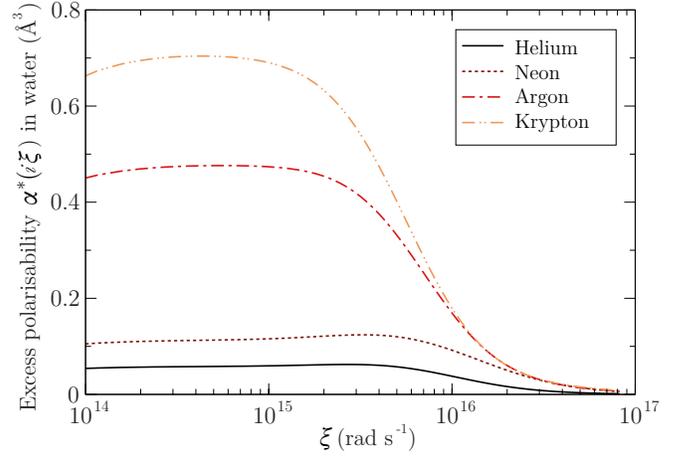}
\caption{(Color online) Excess polarizability of dissolved noble gas atoms in water at imaginary frequencies. The static excess polarizabilities
 do not fit neatly on the graph. Because the static dielectric function of water is so large it gives negative values for the static excess polarizability.
The results (in  \AA$^3$) for He, Ne, Ar, and Kr are:  -0.1375, -0.2655, -1.0045, and -1.7351, respectively.}
\label{figu3}
\end{figure}

\paragraph{Theoretical background.} The calculation of the Casimir--Polder energy for a polarizable particle (such as an ion or atom) embedded in a dielectric medium is standard \,\cite{Maha, Maha2,Spruch,Ninhb} if the dielectric functions and excess polarizabilities for discrete imaginary Matsubara frequencies, 
\be
  i\xi_n=\frac{2\pi ink_\mathrm{B} T}{\hbar}
\ee
are known ($k_\mathrm{B}$ is Boltzmann's constant and $T$ is temperature). 
The permittivity of water and a selection of organic oils is shown in Fig.\,\ref{figu2} for data taken from Ref.~\cite{Masuda}. 
The excess polarizabilities and atomic sizes for the four lightest noble gas atoms were derived as in several papers by Parsons and Ninham \cite{ParsonsNinham2009,ParsonsNinham2010dynpol}, and are presented as functions of $i\xi$ in Fig.\,\ref{figu3}.
Dynamic polarizabilities of noble gas atoms in vacuum were calculated
using \textsc{Molpro} \,\cite{MOLPRO2008} at a coupled cluster singles and
double (CCSD) level of theory. An aug-cc-pV6Z basis set
\cite{WoonDunning1994,PetersonDunning2002} was used for He, Ne and Ar, and
aug-cc-pV5Z was used for Kr \,\cite{WilsonWoonPetersonDunning1999}.
Polarizabilities  $\alpha(i\xi)$ in vacuum were transformed to excess
polarizabilities $\alpha^{*}(i\xi)$  in water via the relation for a dielectric sphere embedded in a dielectric
medium \cite{BoroudjerdiKimEtAl2005,LandauLifshitz-ElectrodynContMedia-v8},  
\be
  \alpha^{*} = R^3  \frac{ \varepsilon_a-\varepsilon_\mathrm{w}  }{ \varepsilon_a + 2\varepsilon_\mathrm{w}  }. 
\ee 
Here $\varepsilon_\mathrm{w}$ is the dielectric function of water and $R$ is the radius of the atom. $ \varepsilon_a$ is the effective dielectric
function of the atomic sphere, estimated  from the atomic polarizability in vacuum as $ \varepsilon_a = 1 + 4\pi\alpha/V$, where
$V$ is the volume of the atomic sphere.

The Casimir--Polder free energy acting on an dissolved atom in water near the interface to a dielectric (oil) with dielectric function $\varepsilon(i \xi_n)$ results from a summation over imaginary--frequency terms\,\cite{Spruch,Maha, Maha2}:
\begin{equation}
F = \sum\limits_{n = 0}^\infty {}^{{}'}  \alpha^*(i \xi_n) g(i \xi_n),
\label{equ1}
\end{equation}
where the prime on the summation mark means the $n=0$ term is taken with half weight, and

\begin{align}
g\left( i\xi \right) =-k_\mathrm{B}T\int\limits_0^\infty  dk k\, \Bigr[r_p \Bigr(\frac{{2 \gamma_\mathrm{w}}}{{\varepsilon_\mathrm{w}}}-\frac{{\xi^2}}{{c^2 \gamma_\mathrm{w}}}\Bigr)-r_s\frac{{ \xi^2}}{{c^2 \gamma_\mathrm{w}}}\Bigr]  e^{-2 \gamma_\mathrm{w} z},
\label{equ2}
\end{align}
is the trace of the dyadic Green's function with Fresnel reflection coefficients for $p$ (TM) and $s$ (TE) modes, respectively,
\begin{equation}
r_p=\frac{ \varepsilon( i\xi) \gamma_\mathrm{w}-\varepsilon_\mathrm{w}( i\xi) \gamma}{\varepsilon( i\xi) \gamma_\mathrm{w}+\varepsilon_\mathrm{w}( i\xi) \gamma}, ~~ r_s=\frac{\gamma_\mathrm{w}-\gamma}{ \gamma_\mathrm{w}+\gamma},
\label{equ4}
\end{equation}
with
\begin{equation}
 \gamma=\sqrt{k^2+\varepsilon(i\xi) \xi^2/c^2},~~~ \gamma_\mathrm{w}=\sqrt{k^2+\varepsilon_\mathrm{w}(i\xi) \xi^2/c^2}.
\label{equ6}
\end{equation}
In the current, nonmagnetic case, the TM mode (reflection coefficient $r_p$) dominates over the TE mode at all separations.

Inspection of Eqs.~\eqref{equ1}-\eqref{equ6} reveals that the sign of the free energy, and hence the force between atom and interface, is governed by the sign of the reflection coefficients $r_s$ and $r_p$. In the non-retarded regime where retardation is slight, the interaction is found by letting $c\to \infty$,
\be
  F_\mathrm{nonret.} \approx  -\frac{k_\mathrm{B}T}{2z^3}\sum\limits_{n = 0}^\infty {}^{{}'} \frac{\alpha^*(i \xi_n)}{\varepsilon_\mathrm{w}(i\xi_n)}\frac{\varepsilon(i\xi_n)-\varepsilon_\mathrm{w}(i\xi_n)}{\varepsilon(i\xi_n)+\varepsilon_\mathrm{w}(i\xi_n)}.
\ee
The sign of the resulting free energy is hence determined by the permittivities of the two media at \emph{all} imaginary frequencies, the sum being cut off by the media and atom becoming transparent at high frequency. 

In the opposite limit, on the contrary, where $z\gg \sqrt{\varepsilon_\mathrm{w}}\xi_1/c$, the exponential factor ensures that the main contribution comes from the term $n=0$, and one obtains an expression depending only on permittivities and polarizability primarily in the quasistatic limit. Thus, while $\alpha^*$ and $r_p$ may take either sign at higher values of $i\xi$, they must both tend to negative values as $i\xi\to 0$, since $\varepsilon_\mathrm{w}$ far exceeds $\varepsilon$ of any oil in this limit. 

Thus the possibility arises for the interaction free energy to change sign as retardation becomes of importance. Although in the extreme large separation limit where \emph{only} the zero Matsubara frequency enters, the interaction is always attractive (the limiting expression may be written out but is somewhat bulky \cite{ellingsen09}), this may occur at very large separations where the Casimir--Polder force is of little interest. As the separation increases, however, the interaction can in principle change sign more than once, as the relative permittivity of water appears larger or smaller than that of oil and atom-cavity, respectively, in the weighted frequency sum.

\begin{figure}
\includegraphics[width=\columnwidth]{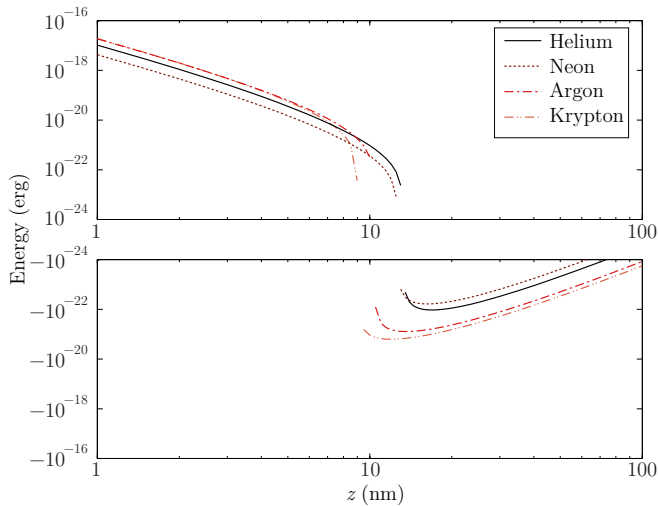}
\caption{(Color online) Retarded Casimir-Polder energy between noble gas atoms dissolved in water near a water-hexadecane interface.}
\label{figu4}
\end{figure}

\paragraph{Numerical investigation.}
We show in Fig.\,\ref{figu4}   the retarded Casimir--Polder energy acting on different noble gas atoms dissolved in water near a water-hexadecane interface. There is a short range repulsion that in the retarded regime  turns attractive. This mimics what was observed for the Casimir--Lifshitz force between  unequal surfaces across a liquid \,\cite{Mund,Milling,Lee,Feiler}.  The Casimir--Polder interaction can in a similar way turn from repulsion to attraction depending on dielectric functions involved. The distance where the Casimir--Polder energy turns attractive depends on the specific atom.

\begin{figure}
\includegraphics[width=\columnwidth]{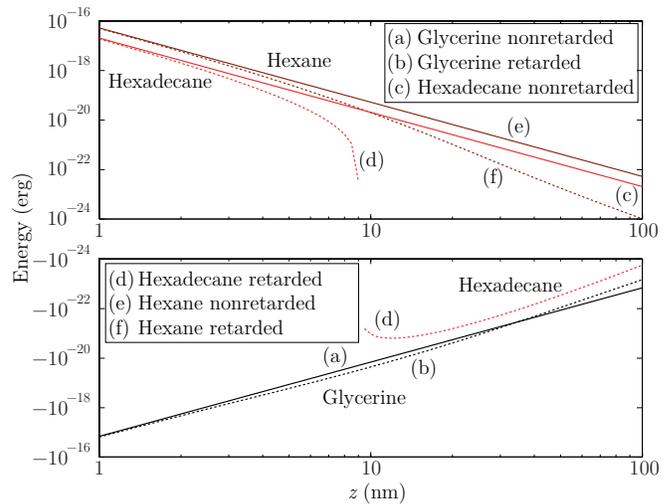}
\caption{(Color online) Retarded and non-retarded Casimir-Polder energy between krypton atoms dissolved in water near different water-oil interfaces. We study in this figure the following oils: hexane, hexadecane, and glycerine.}
\label{figu5}
\end{figure}

In Fig.\,\ref{figu5} both the fully retarded Casimir--Polder energies and the non-retarded van der Waals energies of krypton atoms near different water-oil interfaces are shown. 
Although the difference in permittivity between hexane and glycerine is not great (see Fig.~\ref{figu2}), the atom is repelled from an interface to the former over the full separation range whereas only attraction is seen for the latter.
For hexadecane the non-retarded interaction is repelled at all distances but when retardation is included the interaction above a certain limiting distance turns attractive.

\begin{figure}
\includegraphics[width=\columnwidth]{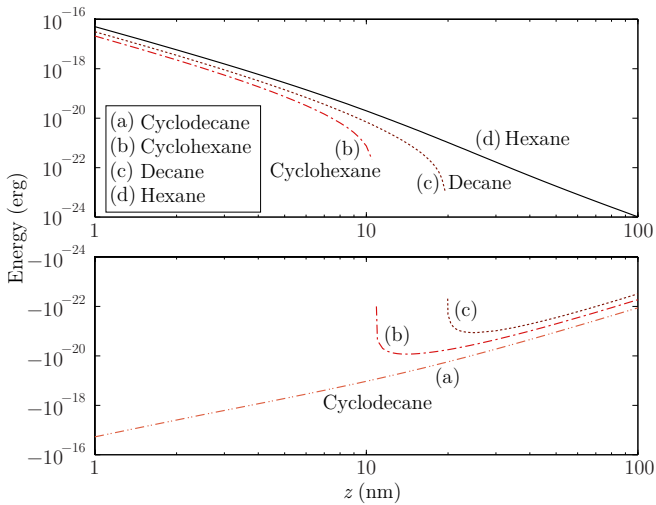}
\caption{(Color online) Retarded  Casimir-Polder energy between krypton atoms dissolved in water near different water-oil interfaces. We study in this figure the following oils: hexane, decane, cyclohexane, and cyclodecane.}
\label{figu6}
\end{figure}

We study  in Fig.\,\ref{figu6}  the retarded Casimir--Polder energy of Krypton atoms near hexane, decane, cyclohexane, and cyclodecane. Here we see the full spectrum from attraction (cyclodecane) to repulsion (hexane) via intermediate oils that have both attractive and repulsive regions (decane and cyclohexane). As can be seen in Fig.\,\ref{figu2} the difference in dielectric functions of these oils are very small, yet these differences are nevertheless sufficient to  produce the very different results observed.

\paragraph{Concluding remarks.} We have seen that the Casimir-Polder force acting on atoms near water-oil interfaces may act to enhance or deplete the amount of atoms near the interface. Depending on the optical properties of atoms, water and oil one can have attraction, repulsion or both acting between a specific atom and oil-water interface. For many oils the effect of retardation is to turn non-retarded van der Waals repulsion to retarded Casimir-Polder attraction.  The usefulness to industrial applications of the results exemplified herein would appear to be immediate. One can envisage separation of different species of particles dissolved in water, for example, by carefully choosing an interface by which one species is attracted, the other repelled, allowing selective adsorption. The same properties would be thought to be useful in biological systems.
From a wider point of view it may be worthwhile to mention that the formation of microsize gas bubbles near a planar surface can be of large technological interest. One typical example of this kind occurs in an aluminum plant, where in the Hall-Heroult cells there are CO$_2$ or CO bubbles near the anode. Phenomenological descriptions of this process under specified conditions can be given using methods from electrodynamics and hydrodynamics, but a detailed knowledge of this kind of phenomena is still lacking. It may well be that when the anode-bubble separations are very small, in the submicron region, the effects predicted in the present paper may come into play. For a recent review dealing with Hall-Heroult cells see, for instance, the work by  Einarsrud and Johansen \,\cite{Einarsrud}.

\begin{acknowledgments}
 MB acknowledge support from a European Science Foundation exchange grant within the activity "New Trends and Applications of the Casimir Effect," through the network CASIMIR.  M.B. and B.E.S. acknowledge support from VR Contract No. 70529001. D.F.P. acknowledges support from the Australian Research Council. We are  grateful to Barry W. Ninham for valuable discussions and to Marin-Slobodan Tomas for finding an error in one of our equations.
\end{acknowledgments}


\begin{thebibliography}{10}
\bibitem{CasPold} H. B. G. Casimir and D. Polder, Proc. K. Ned. Akad. Wet. {\bf 60}, 793 (1948).
\bibitem{london30} F.\ London, Z.\ Phys.\ {\bf 63}, 245 (1930).
\bibitem{Lifshitz} E.~M.\ Lifshitz, {Z}h.\ {E}ksp.\ {T}eor.\ {F}iz.\ {\bf 29}, 94 (1955) [{S}ov.\ {P}hys.\ JETP {\bf 2}, 73 (1956)].
\bibitem{Dzya} I.~E.\ Dzyaloshinskii, E.~M.\ Lifshitz, and  L.~P.\ Pitaevskii, Advan.\ Phys. {\bf 10}, 165 (1961).
\bibitem{barton70} G. Barton, Proc. R. Soc.  A {\bf 320}, 251 (1970).
\bibitem{Ser} Bo E. Sernelius, {\it Surface Modes in Physics} (Wiley-VCH,  Berlin, 2001).
\bibitem{buhmann07} S.~Y.\ Buhmann and D.-G.\ Welsch, Prog.\ Quantum Electron.\ {\bf 31}, 51 (2007).
\bibitem{scheel08} S.~Scheel and S.Y.~Buhmann, Acta Phys. Slov. {\bf 58}, 675 (2008).
\bibitem{AndSab} C. H. Anderson and E. S. Sabisky, Phys. Rev. Lett. {\bf 24}, 1049 (1970).
\bibitem{Rich1} P. Richmond and B. W. Ninham, Solid State Communications {\bf 9}, 1045 (1971).
\bibitem{Haux} F. Hauxwell and R. H. Ottewill, J. Colloid Int. Sci. {\bf 34}, 473 (1970).
\bibitem{Rich} P. Richmond, B. W. Ninham and R. H. Ottewill, J. Colloid Int. Sci. {\bf 45}, 69 (1973).
\bibitem{Masuda} T. Masuda, Y. Matsuki, and T. Shimoda, J. Colloid Int. Sci. {\bf 340}, 298 (2009).
\bibitem{Maha} J. Mahanty and B. W. Ninham, {\it Dispersion Forces}, (Academic Press, London, 1976).
\bibitem{Maha2} J. Mahanty and B. W. Ninham, Faraday Discuss. Chem. Soc. {\bf 59}, 13 (1975).


\bibitem{Elb} M. Elbaum and  M. Schick, Phys. Rev. Lett. {\bf 66}, 1713 (1991).
\bibitem{Wil} L. A. Wilen, J. S. Wettlaufer, M. Elbaum, and  M. Schick, Phys. Rev. B {\bf 52}, 12426 (1995).
\bibitem{Lev} M. Levin, A. P. McCauley, A. W. Rodriguez, M. T. Homer Reid, and S. G. Johnson, Phys. Rev. Lett. {\bf 105}, 090403 (2010).
\bibitem{Mag} M. F. Maghrebi, Phys. Rev. D {\bf 83}, 045004 (2011).

\bibitem{Spruch} F. Zhou and L. Spruch, Phys. Rev. A {\bf 52}, 297  (1995).
\bibitem{Ninhb} B. W. Ninham and P. Lo Nostro, {\it Molecular Forces and Self Assembly in Colloid}, Nano Sciences and Biology, (Cambridge University Press, Cambridge, UK, 2010).
\bibitem{ParsonsNinham2009} D.F. Parsons and B. W. Ninham, J. Phys. Chem. A {\bf 113}, 1141 (2009).
\bibitem{ParsonsNinham2010dynpol} D.F. Parsons and B. W. Ninham, Langmuir {\bf 26}, 1816 (2010).
\bibitem{MOLPRO2008} H.-J. Werner et al., MOLPRO, version 2008.1, a package of ab initio programs, 2008. 
\bibitem{WoonDunning1994} D. E. Woon and T. H. Dunning, Jr., J. Chem. Phys. {\bf 100}, 2975 (1994).
\bibitem{PetersonDunning2002} K. A. Peterson and T. H. Dunning, Jr., J. Chem. Phys. {\bf 117}, 10548 (2002).
\bibitem{WilsonWoonPetersonDunning1999} A. K. Wilson,  D. E. Woon, K. A. Peterson and T. H. Dunning, Jr., J. Chem. Phys. {\bf 110}, 7667 (1999).
\bibitem{BoroudjerdiKimEtAl2005} H. Boroudjerdi, Y.-W. Kim, A. Naji, R.R. Netz, X. Schlagberger and A. Serr, Phys. Rep. {\bf 416}, 129 (2005).
\bibitem{LandauLifshitz-ElectrodynContMedia-v8} L. D. Landau and E. M. Lifshitz, {\it Electrodynamics of Continuous Media}, (Pergamon Press, Oxford, 1960).
\bibitem{ellingsen09} S.~\AA.\ Ellingsen, S.~Y.\ Buhmann, and S.~Scheel, Phys.\ Rev.\ A {\bf 79}, 052903 (2009).
\bibitem{Mund} J. N. Munday, F. Capasso, and V. A. Parsegian, Nature (London), {\bf 457}, 07610 (2009).
\bibitem{Milling} A. Milling, P. Mulvaney, and I. Larson, J. Colloid Interface Sci. {\bf 180}, 460 (1996).
\bibitem{Lee} S. Lee and W. M. Sigmund, J. Colloid Interface Sci. {\bf 243}, 365 (2001).
\bibitem{Feiler} A. A. Feiler, L. Bergstrom, and M. W. Rutland, Langmuir {\bf 24}, 2274 (2008).
\bibitem{Einarsrud} K. E. Einarsrud and S. T. Johansen, Progress in Computational Fluid Dynamics, in press (2012).
\end{thebibliography}
\end{document}